# Twisted terahertz radiation generation using Laguerre-Gaussian laser pulse propagating in axially magnetized plasma

Dinkar Mishra, Saumya Singh, Bhupesh Kumar[1], Pallavi Jha[2]

*Department of Physics, University of Lucknow, Lucknow, Uttar Pradesh, India – 226007*

## Abstract

We present analytical and simulation study of twisted terahertz (THz) radiation generation via propagation of a circularly polarized Laguerre-Gaussian (LG) laser pulse in homogeneous plasma embedded in an axial magnetic field. Analytical formulation is based on perturbation technique and quasi-static approximation. Longitudinal and transverse wakefields generated via laser-plasma interactions are evaluated using Lorentz force and Maxwell's equations in the mildly nonlinear regime. It is observed that two linearly polarized twisted terahertz (THz) radiation beams are generated in mutually perpendicular planes. Superposition of the two beams result in a single linearly polarized twisted THz radiation beam with modified amplitude and polarization direction. Three-dimensional (3-D) particle in cell (PIC) simulations are performed for this configuration using FBPIC code. Graphical comparison of amplitude of the resultant THz beam obtained via analytical and simulation studies is presented.

**Keywords:** laser-plasma interaction, Laguerre-Gaussian laser beams, wakefield generation, radiation generation, twisted THz beam generation, PIC simulation

---

[1] Corresponding Author: Bhupesh Kumar, bhupeshk05@gmail.com
[2] Retired Professor

**Draft 7.0**

## I. Introduction

Interaction of intense, short laser pulses with plasma plays a key role in several applications such as laser wakefield accelerator (LWFA)[1-4], intense radiation sources[5-10], plasma diagnostics[11-13], light sources[14-17], and laser induced nuclear fusion[18-22]. Many viable aspects of laser-plasma interactions have been studied at length in the last decades. Under optimum conditions such interactions may lead to generation of various types of electromagnetic radiations[23-27]. Numerous analytical, simulation and experimental studies have shown that propagation of intense laser pulses when propagated through plasma generate longitudinal as well transverse wakefields. Near infrared to terahertz (THz) range radiation has been observed via propagation of two-color lasers in unmagnetized plasma[28]. Several simulation and numerical studies have shown that terahertz radiation can be generated via interaction of intense laser pulses in magnetized plasma[29-32].

LG laser pulses also known as vortex laser pulses possess several novel properties such as orbital angular momentum (OAM), less diffraction over long distance propagation, versatility in their beam profile including radially symmetric and azimuthally variant intensity patterns. These properties of LG laser pulses find significant role in particle acceleration, radiation generation, and quantum information like intrinsic spin of accelerated particles. Such OAM laser pulses can be easily produced using Gaussian laser pulses with combination of gratings[33], spiral phase plates[34], and spatial light modulators[35-36]. Numerical simulations have shown generation of high intensity attosecond pulse radiation via interaction of LG laser beams with plasma electrons[37]. Simulation and experimental studies by Wang et al. show that twisted THz radiation of 0.3 – 30 THz can be generated by four wave mixing process of two-color LG laser beams propagating in plasma[38].

Propagation of LG laser pulses in magnetized plasma generate donut shaped longitudinal and transverse wakefields oscillating at plasma frequency as reported in an earlier research work. However, the possibility of obtaining twisted THz radiation via transverse wakefield generation in plasma using single LG laser pulse has not been reported so far. Such type of radiation fields may surpass the diffraction limit, leading to super-resolution imaging capabilities. The unique spatial structure of twisted THz radiation allows for selective excitation of specific modes in materials, enabling detailed material characterization. In addition to that, the ability to concentrate energy in specific regions can improve the sensitivity of terahertz spectroscopy for detecting subtle changes in material properties. The high-intensity regions of twisted THz radiation can enhance nonlinear optical effects in materials, leading to more efficient generation of higher harmonics or other nonlinear phenomena.

**Draft 7.0**

The present study deals with an analytical study of twisted THz radiation generation and its validation with quasi-3D PIC simulation results using Fourier-Bessel particle in cell code (FBPIC)[39]. The system comprises of a circularly polarized Laguerre-Gaussian (LG) laser pulse propagating through homogeneous plasma embedded in an external axial magnetic field. Organization of the paper is as follows: Sec. II of the paper contains mathematical formulation of the problem and evaluation of generated longitudinal and transverse wakefields behind the laser pulse. Sec. III of the paper presents discussions on THz generation and its dependence on amplitude of the applied magnetic field and the azimuthal as well as radial order of the LG laser pulse. Sec. IV deals with discussion on simulation results and its comparison with analytical findings. Sec. V of the paper presents summary and conclusions.

## II. Mathematical Formulation

Consider a circularly polarized, LG laser pulse propagating along the $z$-axis in axially magnetized homogeneous plasma having ambient plasma density $n_0$. The pulse profile is considered to be sinusoidal. The electric field component of the propagating laser pulse is given by[40],

$$\vec{E}_{LG} = E_0 \sin \pi \frac{z-ct}{L} \exp\left(-\frac{r^2}{r_0^2}\right) F_g(r,\theta) \left[\hat{x} \cos(k_0 z - \omega_0 t) + \hat{y} \sin(k_0 z - \omega_0 t)\right], \quad (1)$$

where $F_g(r,\theta) = \sqrt{\frac{\rho!}{(l+\rho)!}} \left(\frac{\sqrt{2}r}{r_0}\right)^l F_\rho^l \left(\frac{2r^2}{r_0}\right) e^{il\theta}$, $r^2 = x^2 + y^2$, $\theta = \tan^{-1}\left(\frac{y}{x}\right)$ and $F_\rho^l$, $\rho$, $l$, $r_0$, $\theta$, $k_0$, $\omega_0$, $E_0$, $L$ represent associated Laguerre Gaussian polynomial, radial order, azimuthal order, beam waist radius, azimuthal angle, wave number, frequency, amplitude and pulse length of the circularly polarized LG laser pulse respectively. The external axial magnetic field having magnitude $b_0$ applied along the direction of propagation of the laser pulse is given by,

$$\vec{b} = \hat{z} b_0, \quad (2)$$

Following Ref. 30, Eqs. (1) and (2) are used along with Lorentz force equation to evaluate the expression for the generated longitudinal wakefield. The evaluation method is based on perturbative technique as well as quasi-static approximation (QSA), in the mildly nonlinear regime. The second order differential equation governing lowest order generated longitudinal electric wakefields (oscillating at the plasma frequency) in terms of transformed variables $\xi = z - ct$ and $\tau = t$ is thus given by,

$$\left(\frac{\partial^2}{\partial \xi^2} + k_p^2\right) E_z = -\frac{k_p^2 \varepsilon \pi}{2L(1+\Omega)^2} F_g^2(r,\theta) \sin \frac{2\pi \xi}{L}, \quad (3)$$

where $\varepsilon = mc^2 a_r^2/e$, $a_r^2 = a_0^2 exp(-2r^2/r_0^2)$, $a_0 (= \frac{eE_0}{mcw_0})$. $\Omega \ (= \frac{eb_0}{mcw_0})$ is the cyclotron frequency normalized by frequency of LG laser and $k_p (= \sqrt{4\pi n_0 e^2/mc^2})$ is the plasma wave number. Plasma electrons mass and charge are depicted by $e$ and $m$ respectively while $c$ is the speed of light.

In order to obtain longitudinal wakefields generated behind the laser pulse, Eq. (3) is solved within appropriate limit ($\xi < 0$) for a sinusoidal laser pulse to give Eq. (4) as,

$$E_z = \frac{\varepsilon k_p f}{4(1+\Omega)^2} F_g^{\ 2}(r,\theta)[sink_p\xi + sink_p(L-\xi)], \tag{4}$$

where $F_g^{\ 2}(r,\theta) = \frac{1}{l!}\left(\frac{2r^2}{r_0^2}\right)^l e^{i2l\theta}$ \qquad for $\rho = 0$ and,

$$F_g^{\ 2}(r,\theta) = \frac{1}{(l+1)!}\left(\frac{2r^2}{r_0^2}\right)^l \left(l+1-\frac{2r^2}{r_0^2}\right)^2 e^{i2l\theta} \qquad \text{for } \rho = 1.$$

In order to evaluate generated transverse electric and magnetic wakefields, components of transformed Maxwell's equations are written as,

$$\left(\vec{\nabla}_\perp + \frac{\partial}{\partial\xi}\hat{x}\right) \times \vec{E} = \frac{\partial}{\partial\xi}\vec{B}, \tag{5}$$

and

$$\left(\vec{\nabla}_\perp + \frac{\partial}{\partial\xi}\hat{x}\right) \times \vec{B} = \frac{4\pi}{c}\vec{J} - \frac{\partial}{\partial\xi}\vec{E}, \tag{6}$$

where $\vec{E}$ and $\vec{B}$ represent electric and magnetic fields generated in the wake of the laser pulse and $\vec{J} \ (= -n_0 e\vec{v})$ represents second (lowest) order current density arising due to perturbation of plasma electron velocity. Equations (5) and (6) are solved along with Eq. (4) to give generated transverse components of electric and magnetic wakefields as,

$$E_{x(y)} = -\frac{mc^2}{2e(1+\Omega)^2}\{\frac{f}{2}\left(cosk_p\xi - cosk_p(L-\xi)\right) + sin^2\frac{\pi\xi}{L}\}\frac{\partial}{\partial x(y)}a_r^2 F_g^{\ 2}(r,\theta) + \frac{mc\omega_0\Omega}{4ek_p}\{f\left(sink_p\xi + sink_p(L-\xi)\right)\}\frac{\partial}{\partial y(x)}a_r^2 F_g^{\ 2}(r,\theta), \tag{7}$$

$$B_{y(x)} = -\frac{mc^2}{2e(1+\Omega)^2}sin^2\frac{\pi\xi}{L}\frac{\partial}{\partial x(y)}a_r^2 F_g^{\ 2}(r,\theta) + (-)\frac{mc\omega_0\Omega}{4ek_p}\{f\left(sink_p\xi + sink_p(L-\xi)\right)\}\frac{\partial}{\partial y(x)}a_r^2 F_g^{\ 2}(r,\theta), \tag{8}$$

where $\frac{\partial}{\partial x(y)} a_r^2 F_g{}^2(r,\theta) = (-\frac{4x(y)}{r_0^2}) a_r^2 F_g{}^2(r,\theta) + a_r^2 \frac{e^{i2l\theta}}{l!\,r_0^2} \left\{\left[\frac{2r^2}{r_0^2}\right]^{l-1} 4lx(y)\right\}$ for $\rho = 0$ and,

$\frac{\partial}{\partial x(y)} a_r^2 F_g{}^2(r,\theta) = (-\frac{4x(y)}{r_0^2}) a_r^2 F_g{}^2(r,\theta) + a_r^2 \frac{e^{i2l\theta}}{(l+1)!\,r^2 r_0^4} (r_0^2(l+1) - 2r^2)^2 \left(\frac{2r^2}{r_0^2}\right)^l \left\{2lr_0^2[x(y) -\right.$

$(+)y(x)] - \frac{8x(y)r^2}{r_0^2(l+1)-2r^2}\right\}$ for $\rho = 1$.

Equations (7) and (8) represent transverse electric and magnetic fields propagating in magnetized plasma. It may be noted that the electric and magnetic fields have unequal amplitudes and are generated off-axis while on-axis amplitude is zero.

### III. THz radiation generation

In order to study the possibility of the THz radiation generation, equations (7) and (8) are simplified by substituting $x = 0$ ($y = 0$) to give electric and magnetic wakefields generated in the $y - z$ ($x - z$) plane as,

$E_{x(y)} = +(-)B_{y(x)} = \frac{mc\omega_0\Omega}{4ek_p}\left\{f\left(sink_p\xi + sink_p(L - \xi)\right)\right\} \frac{\partial}{\partial y_{x=0}(x_{y=0})} a_r^2 F_g{}^2(r,\theta),$ (9)

The fields (Eq. 9) are generated behind the sinusoidal laser pulse are maximized ($L \to \lambda_p$) to give,

$E_{xm(ym)} = +(-)B_{ym(xm)} = -\frac{\pi mc\omega_0\Omega}{4ek_p} cosk_p\xi \frac{\partial}{\partial y_{x=0}(x_{y=0})} a_r^2 F_g{}^2(r,\theta),$ (10)

here $\frac{\partial}{\partial y_{x=0}(x_{y=0})} a_r^2 F_g{}^2(r,\theta) = \frac{a_r^2 e^{i2l\theta}}{l!\,r_0^2}\left\{4l\,y(x)\left[\frac{2y^2(x^2)}{r_0^2}\right]^{l-1} - 4y(x)\left(\frac{2y^2(x^2)}{r_0^2}\right)^l\right\}$ for $\rho = 0$ and,

$= \frac{a_r^2 e^{i2l\theta}\left(-\frac{4y(x)}{r_0^2}\right)\left[\frac{2y^2(x^2)}{r_0^2}\right]^l\left(l+1-\frac{2y^2(x^2)}{r_0^2}\right)^2}{(l+1)!} + \frac{a_r^2 e^{i2l\theta}\left[\frac{2y^2(x^2)}{r_0^2}\right]^l\left(l+1-\frac{2y^2(x^2)}{r_0^2}\right)\left[\frac{2l}{y(x)}\left(l+1-\frac{2y^2(x^2)}{r_0^2}\right)-\frac{8y(x)}{r_0^2}\right]}{(l+1)!}$ for $\rho = 1$.

It may be noted from Eq. (10), that the two sets of mutually perpendicular electric and magnetic fields ($E_{xm}, B_{ym}$) and ($E_{ym}, -B_{xm}$) having equal amplitudes are generated. Hence, they constitute two linearly polarized electromagnetic radiation fields oscillating at plasma frequency ($\approx THz$) in two mutually perpendicular planes. The generated radiation fields travelling along the $z$-direction are functions of $x, y$ as well as $\theta$. Therefore, the two radiation fields can be combined to give resultant twisted THz radiation amplitude as $E_{THz}(B_{THz}) = \sqrt{\left[E_{xm}(B_{ym})\right]^2 + \left[E_{ym}(B_{xm})\right]^2}$. The polarization direction of THz radiation will be inclined at an angle $\varphi = \tan^{-1}\{E_{ym}(B_{xm})/ E_{xm}(B_{ym})\}$ with respect to the $x$ axis.

## IV. Simulation

Quasi-3D PIC simulations have been performed with moving window using cylindrically symmetric FBPIC code. Single LG laser pulse was propagated from the origin of the simulation box in to homogeneous plasma embedded in static external magnetic field as discussed in analytical studies. Simulation box having length and breadth equal to 80 $\mu$m and 50 $\mu$m has been divided in to 800 x 50 cells respectively. There are 2 macroparticles per cell along $z$ as well as $r$ direction and 4 macroparticles along $\theta$. Grid spacing was 0.01 $\mu$m in the direction of propagation of the LG laser pulse and 0.1 $\mu$m in the transverse direction. Other parameters considered during the simulation are sufficient enough to avoid any numerical heating which may significantly affect the plasma electrons and the generated wakefield. Pulse duration of the LG laser pulse is 50 fs ($L = 15$ $\mu$m). Courant condition was taken into account while running simulations.

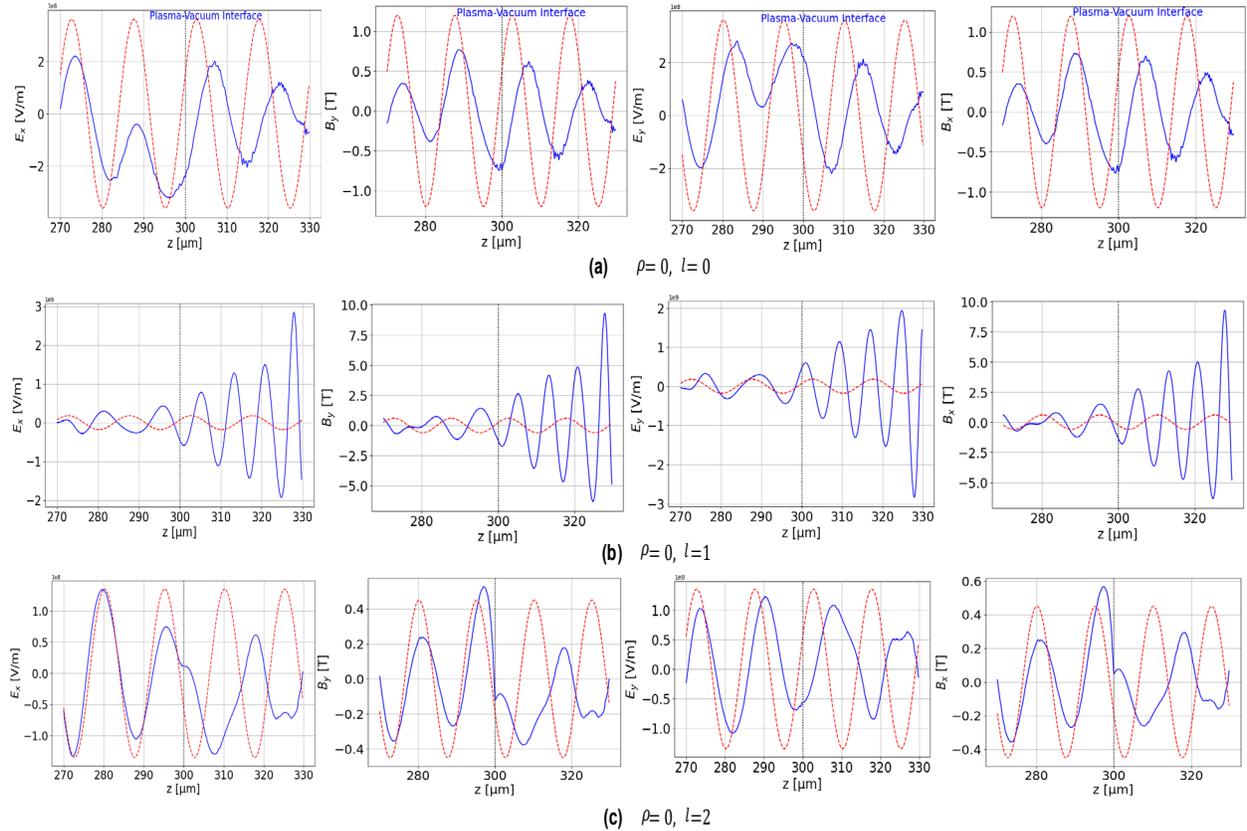

Fig. 1 Analytical (dotted) and simulation (solid) 2-D plots of generated orthogonal components of THz radiation field amplitudes with respect to propagation distance at plasma vacuum interface in magnetized plasma for $a_0 = 0.3, \lambda_p = L = 15\ \mu m, n_0 = 4.958 \times 10^{24}\ m^{-3}, r_0 = 20\ \mu m,\ b_0 = 71\ T$ and off-axis position = 10 $\mu$m where plasma is loaded till 300 $\mu$m.

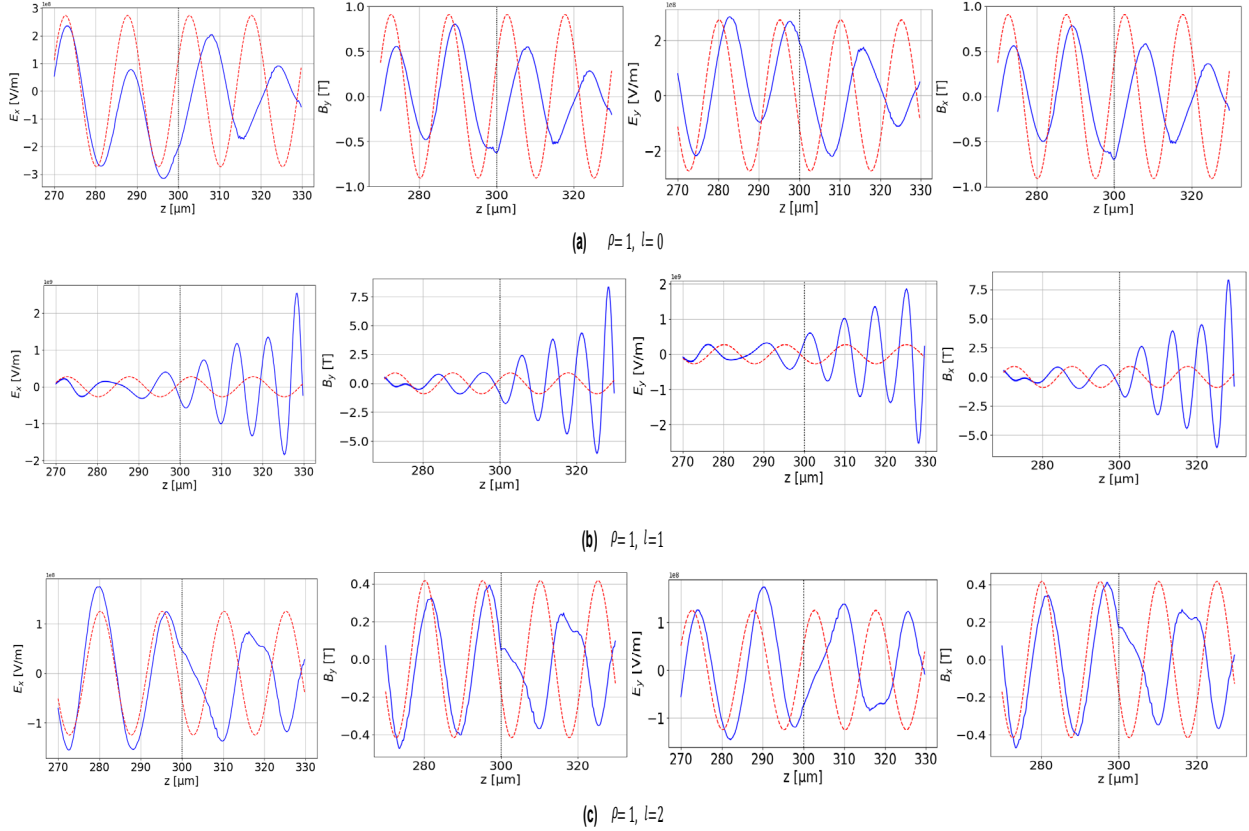

Fig. 2 Analytical (dotted) and simulation (solid) 2-D plots of generated orthogonal components of THz radiation field amplitudes with respect to propagation distance at plasma vacuum interface in magnetized plasma for $a_0 = 0.3, \lambda_p = L = 15\ \mu m, n_0 = 4.958 \times 10^{24}\ m^{-3}, r_0 = 20\ \mu m,\ b_0 = 71\ T$ and off-axis position $= 10\ \mu m$ where plasma is loaded till $300\ \mu m$.

Figures (1) & (2) depict a comparison of components of THz radiation field amplitude obtained at an arbitrary off-axis ($x = y = 10\ \mu m$) position via analytical (eq. 10) as well as simulation studies for $a_0 = 0.3, \lambda_p = L = 15\ \mu m, n_0 = 4.958 \times 10^{24}\ m^{-3}, r_0 = 20\ \mu m,\ b_0 = 71\ T$. Fig. (1) and (2) are plotted for radial order $\rho = 0$ and $\rho = 1$ respectively. Curves (a) and (c) in both figures represent generated THz radiation field amplitude for symmetric (even) azimuthal modes ($l = 0, 2$) where analytical and simulation results are seen to be in good agreement. On the contrary, simulation curves (b) obtained for asymmetric (odd) azimuthal mode ($l = 1$), shows a significant deviation from analytical results where the generated THz radiation field undergoes a modulation in frequency as well as amplitude. Such abrupt behavior (only observed via simulation) may be attributed to the asymmetry factor in odd LG modes which leads to an uneven radial distribution of the electric field, concentrating energy in certain angular directions. This causes a more effective charge separation in the plasma, driving a higher-amplitude wakefield

modulating in frequency as well as amplitude. However, even azimuthal modes (like $l = 0, 2, 4, .....$) have symmetric field distributions around the central axis. This symmetry does not induce the same degree of rotational force in the plasma and therefore does not generate an asymmetric charge separation as strongly as odd modes do.

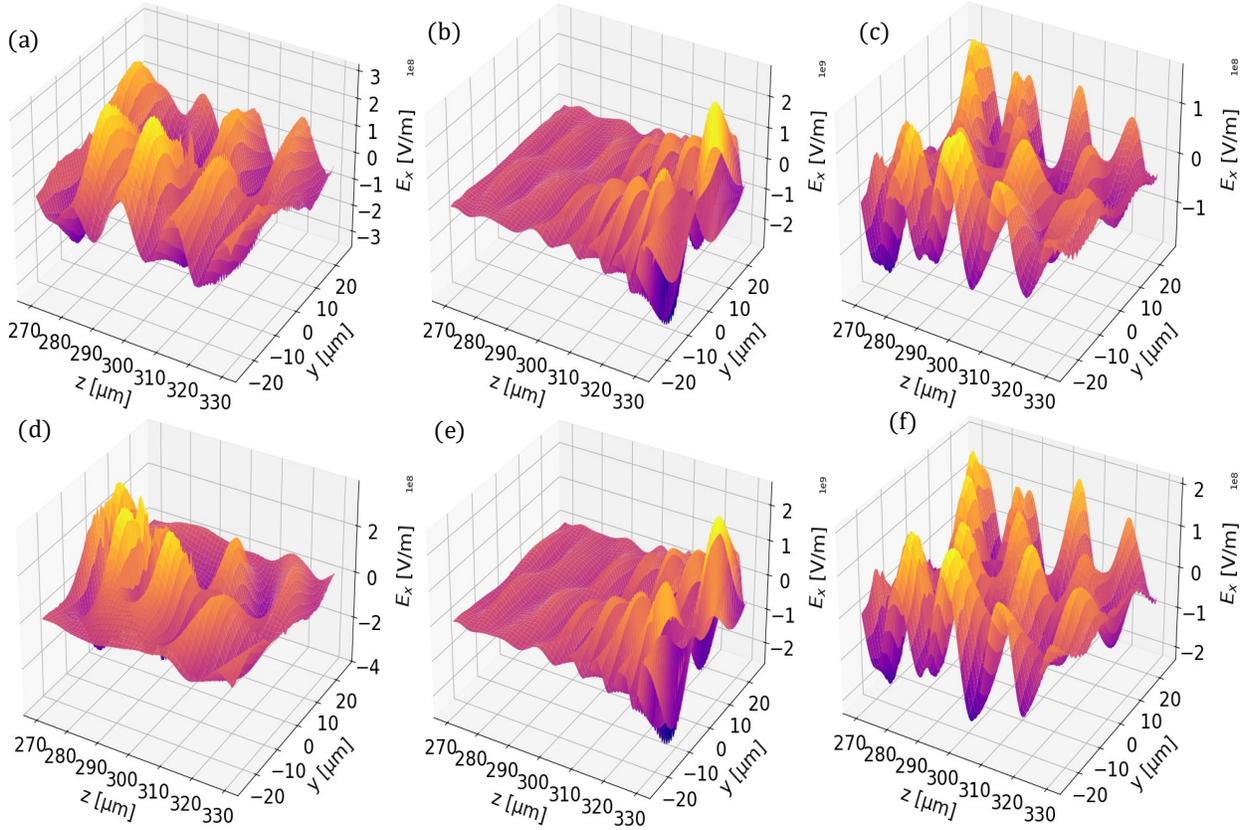

Fig. 3  3-D simulation snapshot of generated THz electric field in $y - z$ plane traversing through plasma vacuum interface at an arbitrary off-axis ($y = 10\ \mu m$) position for (a) $\rho = 0, l = 0$ (b) $\rho = 0, l = 1$ (c) $\rho = 0, l = 2$ (d) $\rho = 1, l = 0$ (e) $\rho = 1, l = 1$ and (f) $\rho = 1, l = 2$ where $a_0 = 0.3, \lambda_p = L = 15\ \mu m, n_0 = 4.958 \times 10^{24}\ m^{-3}, r_0 = 20\ \mu m,\ b_0 = 71\ T$

Figure (3) shows 3-D propagation of generated THz radiation field traversing through plasma vacuum interface in $y - z$ plane for the parameters discussed in Fig.1. It may be noted that, amplitude of the generated THz radiation field is slightly reduced on crossing the plasma vacuum interface due to change in velocity of the generated beam. The difference between amplitudes of the generated THz radiation field components obtained in analytical and simulation results may be attributed to the approximations considered in analytical study.

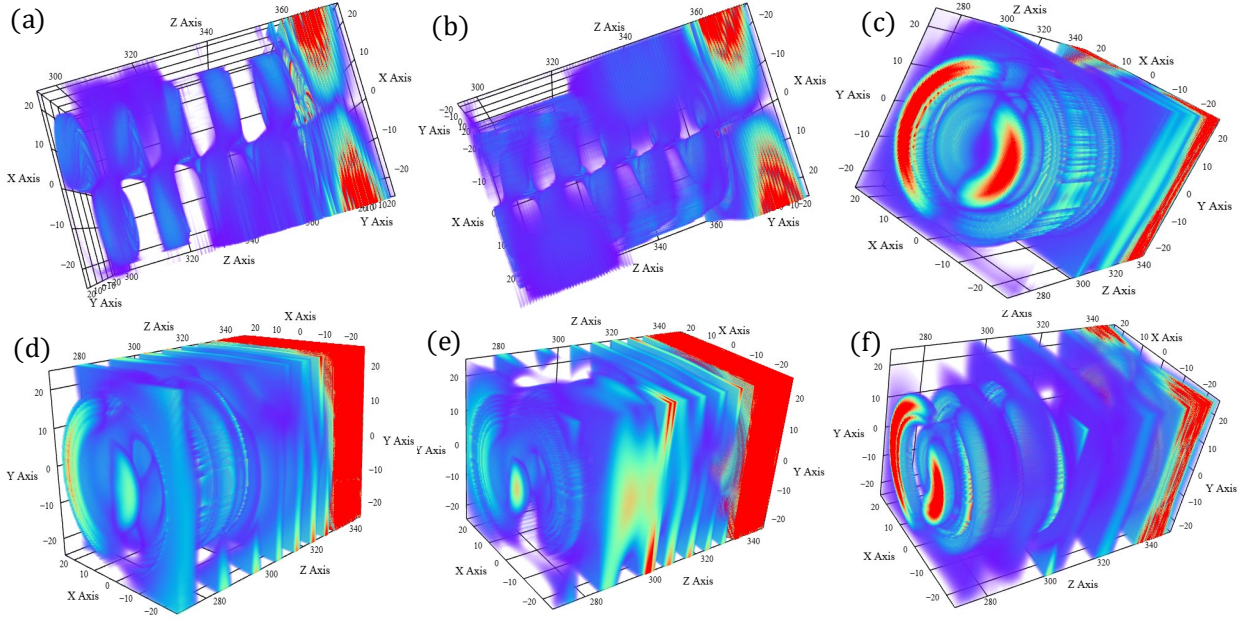

Fig. 4 3-D simulation snapshots depicting resultant linear THz radiation fields (a) $\rho = 0, \; l = 0$ (b) $\rho = 1, \; l = 0$ and twisted THz radiation fields (c) $\rho = 0, \; l = 1$ (d) $\rho = 1, \; l = 1$ (e) $\rho = 0, \; l = 2$ and (f) $\rho = 1, \; l = 2$ for $a_0 = 0.3, \lambda_p = L = 15 \; \mu m, n_0 = 4.958 \times 10^{24} \; m^{-3}, r_0 = 20 \; \mu m, b_0 = 71 \; T$

Figure (4) depicts 3-D simulation screenshots obtained via simulation at the same time step and iteration for $a_0 = 0.3, \lambda_p = L = 15 \; \mu m, n_0 = 4.958 \times 10^{24} \; m^{-3}, r_0 = 20 \; \mu m, b_0 = 71 \; T$. The phase profile of the generated THz radiation field for azimuthal order $l = 0$ ($\rho = 0, 1$) appears to be linear while twisted phase profile is generated for higher azimuthal orders $l = 1, 2$ ($\rho = 0, 1$).

## V. Conclusion

The present paper deals with analytical and simulation study of twisted terahertz (THz) radiation generation via propagation of a circularly polarized Laguerre-Gaussian (LG) laser pulse in homogeneous plasma embedded in a static axial magnetic field. Following Ref. [30], analytical formulation is based on perturbation technique and quasistatic approximation. Wakefields generated via laser-plasma interaction are evaluated using Lorentz force and Maxwell's equations in a mildly nonlinear regime. It is observed that two linearly polarized THz radiation beams are generated in mutually perpendicular planes. Superposition of the two beams result in a single linearly polarized twisted THz radiation beam with modified amplitude and polarization direction. It is also noticed via simulation, that asymmetry in odd LG modes leads to a more effective charge separation resulting in frequency and amplitude modulation of the generated THz radiation.

However, no such modulation is noticed for even LG modes attributed to less effective charge separation as well as symmetric distribution of energy around the central axis. It may be noted that amplitude of the generated twisted THz field is directly proportional to the amplitude of the external static field applied, while its frequency is governed by the plasma density.

The study is significant since it points towards the possibility of generation of intense twisted THz radiation using a single LG laser pulse propagating in plasma. The THz radiation frequency closely matches the plasma frequency and can be tuned via plasma density, while its intensity is governed by the laser strength parameter. Minor discrepancies between analytical and simulation results may be attributed to the approximation scheme used in analytical study. The findings may facilitate detailed material characterization. High-intensity regions within the twisted THz radiation can amplify nonlinear optical effects, leading to more efficient generation of higher harmonics and other nonlinear phenomena.

## Acknowledgments

Dinkar Mishra acknowledges the financial support provided by the Dept. of Science and Technology, Govt of India, under the INSPIRE Fellowship (IF180673) scheme.

## Data Availability

The data that support the findings of this study are available upon reasonable request from the corresponding author.

## Competing Interests

We declare that the authors have no competing interests.